 \documentclass{emulateapj}
\usepackage{epsfig} 
\def\spose#1{\hbox to 0pt{#1\hss}}

\newcommand{\ds}{\displaystyle} \newcommand{\myfrac}[2]{\frac{\ds
    {#1}}{\ds {#2}}}

\let\approxlt=\lesssim   \def\multleft#1{\hbox to size{\vbox
    {\halign {\lft{##}\cr #1}}\hfill}\par} \def\multright#1{\hbox to
  size{\vbox {\halign {\rt{##}\cr #1}}\hfill}\par}
 
\def\boxit#1{\vbox{\hrule\hbox{\vrule\kern3pt\vbox{\kern3pt #1
        \kern3pt}\kern3pt\vrule}\hrule}} 
scaled\magstep2  
scaled\magstep2

\newcommand\aproxgt{\mathrel{%
    \rlap{\raise 0.511ex \hbox{$>$}}{\lower 0.511ex \hbox{$\sim$}}}}
\newcommand\aproxlt{\mathrel{%
    \rlap{\raise 0.511ex \hbox{$<$}}{\lower 0.511ex \hbox{$\sim$}}}}



\begin{document}

\title{Connections Between Local and Global Turbulence in Accretion
  Disks}

\author{Kareem~A.~Sorathia\altaffilmark{1,3},
  Christopher~S.~Reynolds\altaffilmark{2,3},
  Philip~J.~Armitage\altaffilmark{4,5}}

\altaffiltext{1}{Department of Mathematics and Department of
  Astronomy, University of Maryland, College Park, MD 20742-2421}
\altaffiltext{2}{Department of Astronomy and the Maryland Astronomy
  Center for Theory and Computation, University of Maryland, College
  Park, MD 20742-2421} \altaffiltext{3}{Joint Space Science Institute
  (JSI), University of Maryland, College Park, MD 20742-2421}
\altaffiltext{4}{JILA, 440 UCB, University of Colorado, Boulder,
  CO~80309-0440} \altaffiltext{5}{Department of Astrophysical and
  Planetary Sciences, University of Colorado, Boulder, CO~80309-0391}

\begin{abstract}
  We analyze a suite of global magnetohydrodynamic (MHD) accretion
  disk simulations in order to determine whether scaling laws for
  turbulence driven by the magnetorotational instability, discovered
  via local shearing box studies, are globally robust. The simulations
  model geometrically-thin disks with zero net magnetic flux and no
  explicit resistivity or viscosity.  We show that the local Maxwell
  stress is correlated with the self-generated local vertical magnetic
  field in a manner that is similar to that found in local
  simulations. Moreover, local patches of vertical field are strong
  enough to stimulate and control the strength of angular momentum
  transport across much of the disk. We demonstrate the importance of
  magnetic linkages (through the low-density corona) between different
  regions of the disk in determining the local field, and suggest a
  new convergence requirement for global simulations -- the vertical
  extent of the corona must be fully captured and resolved. Finally,
  we examine the temporal convergence of the average stress, and show
  that an initial long-term secular drift in the local flux-stress
  relation dies away on a time scale that is consistent with turbulent
  mixing of the initial magnetic field.
\end{abstract}


\section{Introduction}

The modern theory of accretion disks has been dominated by the
discovery that angular momentum transport can be mediated by
magnetohydrodynamic (MHD) turbulence driven by the magnetorotational
instability (MRI; \citeauthor{bhsurvey} \citeyear{mri1},
\citeyear{bhsurvey}).  Although analytic treatments of the MRI suffice
to establish that the instability exists and rapidly develops toward
turbulence, numerical work has been at the forefront of the effort to
characterize the resulting angular momentum transport. Simulations of
the MRI require difficult compromises because arguably important
scales span a wide range between the global scale $\lambda \sim r$, an
intermediate scale $\lambda \sim h$ (where $h$ is the disk scale
height) that roughly defines the scale over which shear dominates
turbulent fluctuations, and viscous and resistive dissipation scales
$\lambda_\nu$ and $\lambda_\eta$. As in many other astrophysical
problems, the latter are usually so small that in the disks under
consideration it is impossible to run simulations that capture the
physical dissipative scales.

To date, the bulk of our numerical understanding of the MRI has been
derived from local shearing-box simulations
\citep{hawley95,brandenburg95}.  In a local shearing-box model, one
studies the local dynamics of an orbiting patch of the accretion flow
using a (co-rotating) Cartesian coordinate system by including
Coriolis forces and shearing boundary conditions. Since only a small
patch of the disk is being simulated, this approach maximizes the
separation between the intermediate driving scale of the turbulence
and the dissipative scales.  For our purposes, the result that in
local simulations the saturation level of magnetic fields and the
strength of angular momentum transport are found to scale with the
vertical flux threading the simulation domain
\citep{hawley95,sano04,boxscaling} will be of particular interest.

Local simulations have known limitations \citep{regev08,bodo08}. By
construction, they enforce the local conservation of quantities (in
particular the net magnetic flux) that in reality are only globally
conserved. In many older implementations they also enforce periodicity
(in radius, azimuth, and in some instances also height) on a scale
that may be small enough to impact the results. Recently, a number of
authors \citep{davis09,guan09,johansen09} have used larger than usual
shearing-box simulations to quantify whether these limitations matter
for practical purposes. In this paper we address the same problem from
the other direction. We analyze small patches of {\em global} disk
simulations in an attempt to determine whether the disk behaves as if
it were a collection of shearing-boxes. Our specific goal is to
ascertain whether the relationship between local $r\phi$-component of
the magnetic stress and vertical magnetic flux that is found in local
simulations \citep{boxscaling} is recovered in local patches of global
simulations.   As a result, we uncover the importance of the magnetic
connectivity of the disk and the need to fully capture the vertical
extent of the corona.    

This paper is organized as follows. \S2 briefly describes the global
simulations that we employ. \S3.1 describes the vertical structure of
the disks we simulate with an emphasis on the distribution of magnetic
flux.  We discuss, in Section~\S3.2, the results of our study of
instantaneous correlations between magnetic flux and stress. The
flux-stress connection is explored in more detail in Section~\S3.3
(where we discuss the nature of the transition point in the
flux-stress relation) and Section~\S3.4 (where we examine the temporal
correlation of stress and vertical flux in co-moving patches).
Sections~\S3.5 and \S3.6 describe the dependence of angular momentum
transport on vertical domain size and time respectively.  Section~\S4
presents our conclusions.

\section{Global Simulations of Thin Accretion Disk}

The simulations used here are very similar to those described by
\citet{christhin}.  We use the ZEUS-MP code \citep{zeus2d1,hayes06} to
solve the equations of ideal MHD in three dimensions.  We modified the
basic version of the code to incorporate a Paczynski-Wiita
pseudo-Newtonian gravitational potential (as a first approximation of
the gravitational field about a Schwarzschild black hole;
\citeauthor{postnewton} \citeyear{postnewton}) and performed the
simulation in cylindrical polar coordinates. Our simulations are ideal
MHD in the sense that no explicit resistive or viscous dissipation is
included; all dissipative processes are due to the discretization of
the spatial domain and hence occur close to the grid scale.
Furthermore, we integrate an internal energy equation assuming an
adiabatic equation of state with $\gamma=5/3$.  Energy is lost from
the domain when magnetic fields undergo numerical reconnection.

The initial disk is in a state of Keplerian rotation (with respect to
the pseudo-Newtonian potential), and is in vertical hydrostatic
equilibrium with a constant scale-height $h$.  Thus, the initial
density, pressure and velocity field is
\begin{equation}
  \rho(r,z) = \rho_0(r) \exp(-\myfrac{z^2}{2h^2}),
\end{equation}
\begin{equation}
  p(r,z) = \myfrac{GMh^2}{(R-2r_g)^2R}\rho(r,z),
\end{equation}
\begin{equation}
  v_\phi = r\Omega = \myfrac{\sqrt{GMr}}{r-2r_g},\hspace{1cm}v_z=v_r=0,
\end{equation}
where $r$ represents the cylindrical radius, $R = \sqrt{r^2+z^2}$,
$r_g = GM / c^2$ and $h = 0.05r_{isco} = 0.3r_g$.  We set the initial
midplane density to be $\rho_0(r)=1$ beyond the innermost stable
circular orbit (ISCO) at $r_{isco}=6r_g$, and $\rho_0(r)=0$ within the
ISCO. The radial boundary conditions correspond to zero-gradient
outflow, while periodic boundaries were imposed on both the $\phi$ and
$z$-boundaries (this last choice, which is made in order to avoid
field-line snapping and other numerical problems, is discussed further
in \citeauthor{christhin} \citeyear{christhin}).

The initial magnetic field is specified in terms of a vector potential
of the form,
\begin{eqnarray}
  A_\phi &=& A_0f(r,z)p^{1/2}\sin(\myfrac{2\pi r}{5h}), \\
  A_r & = & A_z = 0.
\end{eqnarray}
Here $f$ is an envelope function that is unity in the disk body and
smoothly goes to zero away from the main body so as to avoid
unphysical interactions with the boundaries.  This results in a
magnetic field topology consisting of distinct poloidal field loops of
alternating orientation throughout the main body of the simulation.
Of importance to the current discussion is that there is no net
vertical magnetic flux threading the disk as a whole.  The constant
$A_0$ is chosen to ensure that the magnetic field strength is
normalized so that the ratio of volume-integrated gas and magnetic
pressure $\beta\approx 10^3$.

A set of simulations were run to span a range of numerical parameters,
specifically varying the vertical and radial resolution, as well as
the vertical and azimuthal extent of the domain. A comparison of
simulations with varying azimuthal extents suggest a negligible
dependence on this parameter, and as such all the simulations
considered here use the same $30^\circ$ wedge-shaped azimuthal
domain. The vertical domain size was found to be important, and we
will discuss the role of this parameter later in this paper. A
detailed study of the dependence on resolution, requiring much greater
computational expense, is deferred to a later work.  All simulations
presented here have a radial domain $r\in(4r_g,16r_g)$.  Details of
the simulations considered are given below in Table 1.

\begin{table*}
\centering
  \begin{tabular}{|c||c|c|c|c|c|}
    \hline
    Run ID & Resolution & Vertical Extent & Total Orbits & Mean $\alpha_M$ & Standard Deviation\\ 
    & (R,$\phi$,z)  & (in $r_g$) & (at $r_{isco}$) & & of $\alpha_M$\\
    \hline
    Thin.M-Res.12z & (240,32,512)  & 12 & 122 & 0.008 & 0.0011\\
    Thin.M-Res.6z & (240,32,256) & 6 & 664 & 0.0086 & 0.0016\\
    Thin.M-Res.3z & (240,32,128) & 3 & 112 & 0.0061 & 0.001\\
    \hline
  \end{tabular}
  \caption{The spatial resolution, vertical domain size, duration of the simulations 
    analyzed in this work.  Also included are the mean and standard deviation of $\alpha_M$, defined by equation~\ref{alpha}, between 50 and 100 orbits at $r_{isco}$.} 
\end{table*}

\section{Results}

\subsection{Vertical Structure of Thin Disks}

Our primary goal is to study the instantaneous correlation between
stress and magnetic flux within the simulated disk.  In a stratified
disk this correlation may vary with height above the disk midplane,
and hence we start by considering how the mean magnetic field
structure varies vertically in our simulations.
 
To analyze the simulations the principle quantity of interest is the
$r-\phi$ component of the Maxwell stress tensor,
\begin{equation}
  M_{r\phi} = \frac{B_r B_\phi}{4 \pi},
\end{equation}
which dominates MRI-driven angular momentum transport
\citep{bhsurvey}.  In a turbulent disk, both $M_{r\phi}$ and other
physical quantities of interest are complicated functions of space and
time.  To make sense of them we use temporal and spatial averages. Run
{\tt Thin.M-Res.6z} has the longest duration of any of our
simulations, and we use this run to construct a representative
vertical profile of the magnetic structure of the disk.  To reduce the
effects of spatial intermittency in the turbulence, we azimuthally
average over the entire domain and average over a small radial range
centered about a fiducial radius in the body of the disk ( $8r_g - h <
r < 8r_g + h$) .  To smooth out the temporal variability and isolate
the behavior of the disk in a saturated turbulent state we time
average over 400 ISCO orbits starting at orbit 50.

\begin{figure}
  \includegraphics[width=0.45\textwidth]{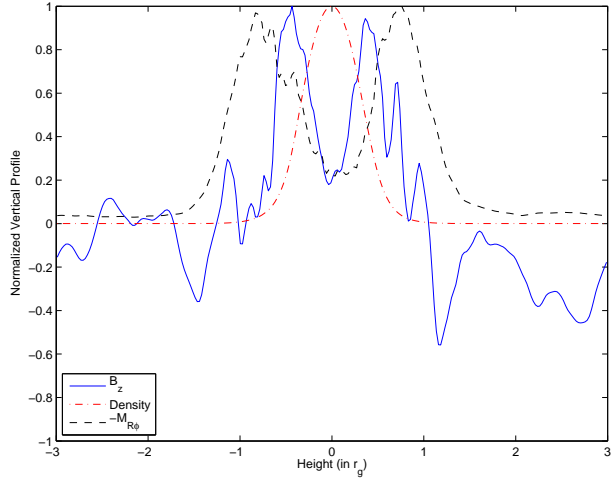}
  \caption{Vertical structure of physical quantities averaged over
    four hundred orbits for run {\tt Thin.M-Res.6z}.}
  \label{vertstruct}
\end{figure}

The results are given in Figure~\ref{vertstruct}, which shows the
vertical profiles of the relevant quantities scaled to their maximum
values.  Our interest in vertical structure is predominantly in the
magnetic fields, and in particular the vertical magnetic flux, $B_z$,
and the magnetic stress, $M_{r\phi}$.  However, we also plot the
density, $\rho$, to highlight the contrast between the relatively
unmagnetized midplane of the disk and the sparse magnetized ``corona"
away from the midplane.  The obvious reason for the formation of these
two disparate regions is magnetic buoyancy resulting from the effect
of vertical gravity, but this may be overly simplistic. Also of
interest is the double-peak vertical profile of vertical flux and
stress.  Broadly similar results are seen in a subset of the
stratified local simulations of \cite{miller00} and a subset of the
global simulations of \cite{fromang06}.  It is interesting that the
vertical location of the peak field and stress seem to approach constant values 
rather than growing monotonically with time.  Whether
the region of strong flux is trapped, possibly due to magnetic tension
from field lines connecting it to the midplane, or is continually
dissipating while outflowing, is currently unclear.

The local flux-stress relationship we will compare our results to is
based on unstratified shearing box simulations and the rich vertical
magnetic structure due to stratification in our global disk means that
there is necessarily some ambiguity in the comparison. In what
follows, we therefore analyze the flux-stress relationship within the
global simulation not just at the midplane but also as a function of
height. One should note that the offset of the peak vertical flux and
stress from $z=0$ means that high values of flux and stress are only
accessible away from the midplane.

\subsection{A local flux-stress correlation in a global disk}
\label{sec:fluxstress}

Although our simulations have zero net vertical flux, small patches of
the disk {\em are} instantaneously threaded by a vertical field. We
seek to determine whether the Maxwell stress tracks this transient
vertical field in the same way as it would in a local simulation where
the vertical field is persistent \citep{hawley95,boxscaling}. To
proceed, we break up the global simulation domain at each timestep
into several hundred small cylindrical wedges of size $\Delta z =
\Delta r = h$ and $\Delta \phi \approx 0.1$.  Within each wedge we
average to obtain a single estimate of the magnetic stress
($M_{r\phi}/p$, normalized to the local gas pressure) and the local
vertical flux, which we express in terms of the wavelength of the most
unstable MRI mode,
\begin{equation}
  \lambda_{\rm MRI} = 2 \pi \left( \frac{16}{15} \right)^{1/2} \frac{{\bar v}_{Az}}{\Omega_0},
\end{equation}
where $v_{Az} = B_z / \sqrt{4 \pi \rho}$ is the Alfven speed
corresponding to the vertical field component. Note that because we
scale the stress by the local pressure (a decreasing function of
height), we immediately introduce a height dependence.  Choosing
instead to scale by the midplane pressure (only a function of radius)
still yields a height dependence and as such we are confident that the
height dependence seen is not solely a consequence of the pressure
scaling.

To avoid early transients, our analysis excludes the period prior to
the first 50 ISCO orbital periods. To improve our statistics (and to
give a measure of the convergence of our results given the finite
duration of the run) we consider wedges that are centered at $z=\pm
[0,1,2,3] h$ and plot results for the samples separately. The
resulting pairs of flux-stress values from all of the wedges and all
of the snapshots in time were binned according to (logarithmic)
vertical flux in order to diagnose trends.

\begin{figure}
  \includegraphics[width=0.45\textwidth]{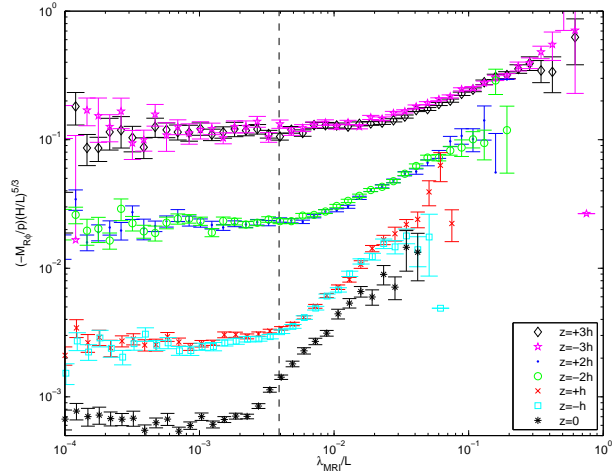}
  \caption{Flux versus stress averaged over four hundred orbits for
    run {\tt Thin.M-Res.6z}.}
  \label{longfvs}
\end{figure}

The resulting flux-stress relations for all the simulations considered
are broadly similar.  Our best statistics come from the long duration
run {\tt Thin.M-Res.6z}, and the flux-stress relation for this case is
plotted in Figure~\ref{longfvs}.  The stress is observed to be flat
for weak vertical fields (small $\lambda_{\rm MRI}$), while for larger
field strengths we have approximately $M_{r \phi} \propto \lambda_{\rm
  MRI}$.  This may be compared with the local scaling relation derived
from unstratified simulations, which can be written in the form
\citep{boxscaling},
\begin{equation}
  \label{mriscale}
  \myfrac{-M_{r\phi}}{P}\Big(\myfrac{H}{L}\Big)^{(5/3)} = 0.61 \times \left\{
    \begin{array}{ll}
      \Delta / L & : \lambda_{MRI} \leq \Delta ,\\
      \lambda_{MRI}/L & : \Delta < \lambda_{MRI} \leq L, \\
      0 & : L < \lambda_{MRI} ,
    \end{array}
  \right.
\end{equation} 
where $\Delta$ and $L$ are the vertical grid cell size and total box
size respectively.  The pressure scale height $H$ is given by

\begin{equation}
  H = \left( \frac{2}{\gamma} \right)^{1/2} \frac{c_s}{\Omega_0},
\end{equation}
where $\Omega_0$ is the Keplerian angular velocity and $c_s$ the sound
speed.  Note the distinction between the locally defined $H$ and the
globally constant $h$.  In comparing our results to those obtained for
local simulations, we consider $L$ to be the size of the wedge and $H$
to be the locally defined pressure scale height.
 
Our simulations do not spontaneously develop vertical fields strong
enough to quench the MRI (and hence we do not sample the $\lambda_{\rm
  MRI} > L$ regime), but the behavior of patches at low and
intermediate vertical fluxes is qualitatively the same as that found
in local simulations.  In contrast to the local (unstratified) results
of \citet{boxscaling} is the strong dependence on height in our
stratified simulations.  In addition to the fact that the largest
values of flux are only accessible at large heights is the fact that
the stress response to flux is also height dependent.  There is a
height-dependence of the transition point between low and intermediate
flux, as well as in the slope of the intermediate flux regime.  Of
particular note is the location of the transition point itself.  The
vertical line marked in Figure~\ref{longfvs} is given by $\lambda_{\rm
  MRI}=\Delta/20$, and approximately marks the location of the
transition point at $z=\pm H$.  This stands in contrast to the
transition point for local unstratified simulations, $\lambda_{\rm
  MRI} = \Delta_z$.  We return to a discussion of the physics of this
transition in Section~\ref{sec:transition}.

\begin{figure}
  \includegraphics[width=0.45\textwidth]{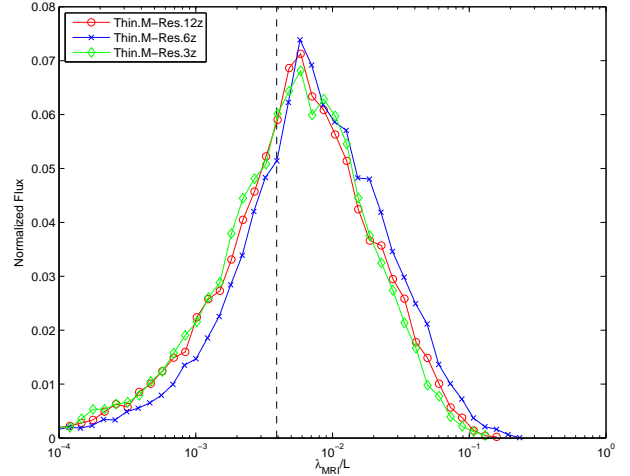}
  \caption{Normalized flux distributions for all simulations, computed
    between 50-100 ISCO orbits. The dashed vertical line indicates,
    approximately, the strength of vertical field above which magnetic
    stress in the disk is stimulated.}
  \label{fluxcomp}
\end{figure}

The fact that transient self-generated vertical flux is able to
stimulate the local stress in the same manner as occurs in local
models is primarily a formal result, although it does lend some
credence to models in which patches of vertical field are assumed to
have a physical identity \citep{spruit05}. Of greater import is the
observation that, across much of the disk, the self-generated field is
strong enough to fall into the linear regime of the flux-stress
relation. Figure~\ref{fluxcomp} shows the distribution of flux through
patches in all three. The vertical flux distributions are
approximately symmetric in log-space.  We find that, at any instant in
time, about half of the area of the disk is threaded by a field strong
enough to control the stress. We interpret this to mean that in a
zero-net field global simulation, much of the disk sees a field strong
enough to control its dynamics.  In other words, the dynamics of the
disk is strongly influenced by the connectivity of the self-generated
magnetic field between different patches of the disk.

In these results we again see the importance of the magnetized region
away from the midplane.  Not only is this the location of the largest
values of flux, and thus stress, but for the same flux the stress
response is higher.  This increased stress response to flux suggests
that the corona is important not just as a warehouse of magnetic
energy, but also has the ability to use this magnetic energy more
efficiently to induce angular momentum transport.  One possible
explanation for this efficiency is the ability of the corona to
mediate magnetic links through radially disparate regions of the disk
due to the presence of coronal field loops.  In
Section~\ref{sec:vertical_domain} we consider the effect of vertical
domain size on the saturated turbulent stress as a means of diagnosing
whether truncating the corona affects the dynamics of a disk.  Before
that, however, we discuss the nature of the transition point in the
flux-stress relation as well as study the connection between the local
vertical magnetic fields and stresses via their temporal correlation.

\subsection{The nature of the transition point in the flux-stress
  relation}
\label{sec:transition}

To reiterate, the flux-stress relation obtained from our global
simulation shows a transition at approximately
$\lambda_{MRI}\sim\Delta/20$, in contrast with $\lambda_{MRI}\sim\Delta$
found from local unstratified simulations \citep{boxscaling}. 
Neither of these transitions can correspond straightforwardly to the condition
that the fastest growing MRI mode is resolved.   To resolve a mode in a
ZEUS-like scheme requires that the wavelength is spanned by at least
$\sim 8$ computational zones.   Since $\lambda_{MRI}$ is, by
construction, the wavelength of the fastest growing mode corresponding
to the net vertical magnetic field, the condition that the fastest
growing MRI mode is resolved corresponds to $\lambda_{MRI}\sim
8\Delta$. Our result then implies that magnetic fields that are 
very weak -- in the sense that their fastest growing mode cannot 
be resolved -- nonetheless have an important influence on the 
dynamics of our global disk.

We do not have a quantitative explanation of why $\lambda_{MRI}\sim\Delta/20$. 
On general grounds, however, we note that we would expect that 
the transition point would lie at $\lambda_{MRI}\ll 8\Delta$. A given 
vertical field is unstable not just to the fastest growing mode, but 
also to a whole spectrum of slower-growing modes that have longer 
wavelengths that are more easily resolvable numerically. Plausibly, 
the transition point will then correspond to the condition that we resolve
the slowest growing mode that grows appreciably before it is truncated
by non-linear coupling to other MRI modes or some other aspect of the
physics (e.g. a dynamo cycle). If this is the case, then it is unsurprising 
that the transition point varies between local and global simulations, 
since the time scale available for a mode to grow may well depend 
on the presence or absence of a low density disk corona within 
which the MRI is not active.

To consider this more quantitatively, consider purely vertical MRI
modes ($k_r=k_\phi=0$) in a thin accretion disk (so that radial
gradients of pressure and entropy can be neglected).  Let the
spacetime dependence of the modes be $e^{i(\omega t-kz)}$. The
dispersion relation for these modes \citep{mri1} reads
\begin{equation}
\tilde{\omega}^4-\kappa^2\tilde{\omega}^2-4\Omega_0^2 k^2v_A^2=0,
\end{equation}
where $\tilde{\omega}^2=\omega^2-k^2v_A^2$ and $\kappa$ is the radial
epicyclic (angular) frequency.  We wish to examine modes with
wavelengths much longer than the fastest growing mode, i.e., with
$|kv_A/\Omega_0|\ll 1$.  Rewriting in terms of the growth rate,
$\sigma=-i\omega$ and expanding the dispersion relation to lowest
order in $k^2v_A^2/\Omega_0^2$ gives
\begin{equation}
\sigma^2=\left[4\left(\frac{\Omega_0}{\kappa}\right)^2 -1\right]k^2v_A^2.
\end{equation}
Suppose that a given mode can grow exponentially for a time $\tau$
before it is truncated by mode coupling or some other unspecified
physical process.  Then, the slowest growing mode that actually
experiences significant growth (hereafter, the slowest appreciably
growing mode [SAGM]) has $\sigma_{sagm}=2\pi/\tau$ and a wavenumber
given by
\begin{equation}
k_{sagm}^2v_A^2=\frac{4\pi^2\kappa^2}{\tau^2(4\Omega_0^2-\kappa^2)}.
\end{equation}
Our hypothesis is that the transition point in the flux-stress
relation corresponds to the point where the slowest appreciably
growing mode is just resolvable, i.e., where $\lambda_{sagm}\equiv
2\pi/k_{sagm}\sim 8\Delta$.   This predicts a transition point at
\begin{equation}
\lambda_{MRI}\sim \frac{16\kappa}{15^{1.2}\pi
  (4\Omega_0^2-\kappa^2)^{1/2}}\, \frac{\Delta}{\tau^\prime}, 
\label{eq:lmri}
\end{equation}
where $\tau^\prime\equiv\tau/t_{orb}$, $t_{orb}$ being the orbital
period at that radius.   

Equation~\ref{eq:lmri} offers some insight into the transition point
found in the flux-stress relations that we have been considering.  In
the local unstratified simulations of \citet{boxscaling}, the implicit
potential is Newtonian ($\kappa^2=\Omega_0^2$) and we find that
$\lambda_{MRI}\sim \Delta$, implying $\tau\sim t_{orb}$.  In the
global simulations presented here, we find the transition point at
$\lambda_{MRI}\sim \Delta/20$ which (accounting for the fact that
$\kappa^2<\Omega_0^2$ in the pseudo-Newtonian potential) gives
$\tau\sim 5-10t_{orb}$.  Thus, within the framework of this argument,
the difference in the location of the transition point between the
local unstratified and the global simulations is due to a difference
in the robustness of the long wavelength and slowly growing modes;
slowing growing modes appear to be able to grow for longer within the
global simulation before being truncated.  The nature of this
difference, which must be closely related to the saturation of the
turbulent state, is beyond the scope of this paper and will be
explored in future work.

\subsection{Temporal correlations in flux-stress}
\label{sec:temporal}

The results of \S\ref{sec:fluxstress} suggests that the (fluctuating)
magnetic flux threading a local patch of the disk determines the
$r-\phi$ component of the magnetic stress generated by the turbulence
in that patch.  If the vertical magnetic flux is indeed the causal
agent in determining the stress, we expect a temporal lag between
fluctuations in the magnetic flux and the resulting variations in the
stress.  On the basis of experiments with local simulations
\citep{hawley96}, we expect this lag to be approximately two (local)
orbital periods.  Thus, we expect the temporal lag to increase with
radius in the disk due to the increasing orbital period.

To search for this lag, we use {\tt Thin.M-Res.6z} and output the 3-d
structure of the disk once every 0.1 ISCO orbits during the interval
between 50--90 ISCO orbits (this is 10 times the nominal data output
rate).  Using these 400 snapshots of the disk structure, we then
computed the instantaneous vertical magnetic fluxes and magnetic
$r-\phi$ stresses in families of co-moving wedges at three radii $r\in
\{8r_g, 10r_g, 12r_g\}$.  The azimuthally averaged value of $v_\phi$
at each radius was used to track a given comoving wedge between
timesteps. This procedure is not fully lagrangian, because it does not
account for the radial movement or fluctuating azimuthal velocity of a
comoving patch, but we expect these effects to be negligible for the
short timeframe under consideration.  The time-series of magnetic flux
and stress for each wedge were then cross-correlated and, finally, the
cross-correlations for all wedges at a given radius were averaged.

\begin{figure}[t]
  \includegraphics[width=0.45\textwidth]{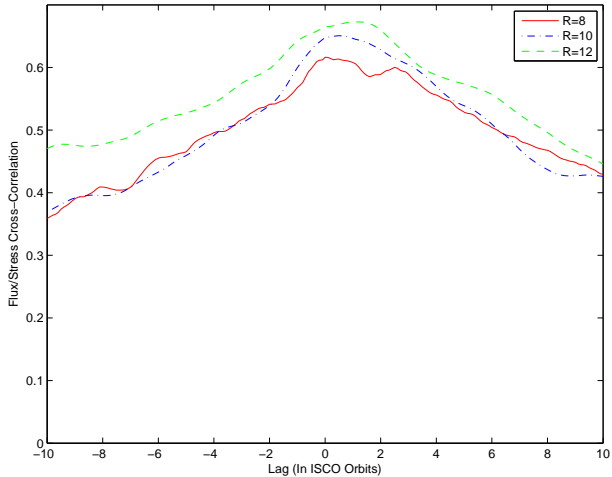}
  \caption{Cross-correlation of flux and stress evaluated in co-moving
    patches of the disk.  Positive lags imply that changes in the
    vertical flux occur, on average, prior to changes in the magnetic
    stress.}
  \label{crosscorr}
\end{figure}

The resulting averaged temporal cross-correlations are shown in
Fig.~\ref{crosscorr}.  At each radius we see a strong instantaneous
correlation, likely due to the immediate shearing of perturbed
vertical fields.  However, in general, the cross-correlation is biased
toward positive lag.  This is consistent with what we would expect,
namely that the presence of vertical flux will feed the MRI and result
in enhanced transport.  Of note is the fact that the inner-most radius
considered, $R=8r_g$ exhibits a double peak structure whereas this is
unresolved at higher radii.  Also peculiar is the fact that the
outer-most radius, $R=12r_g$ is significantly less biased towards
positive lag than the other radii under consideration.  A further
exploration of these issues is beyond the scope of this paper, and
will be explored in future work employing orbital advection algorithms
and test-particle tracers in order to correctly follow the evolution
of a local patch.

\subsection{The dependence of stress on vertical domain size}
\label{sec:vertical_domain}

When considering the correlation between vertical flux and stress we
work directly with $M_{r\phi}$, but when studying possible trends in
average stress with domain size we instead define an effective
$\alpha$-parameter \citep{shakura73}.  For each snapshot we define a
density-weighted spatial average via,
\begin{equation}
  \label{alpha}
  \alpha_{\rm M} = \Bigg\langle \myfrac{\int -\myfrac{\rho M_{r\phi}}{p}dz}{\int \rho dz} \Bigg\rangle_{\phi,r \in (7r_g,12r_g)}.
\end{equation} 
The restriction on the radial range of the averaging is designed to
ignore the plunging region of the accretion flow ($r\approxlt 6r_g$)
and any effects related to the outer radial boundary.  Density
weighting is used in the vertical direction to take into account the
low density, highly magnetized regions while still allowing the
dominant contribution to the integral to come from the denser
mid-plane of the disk.

\begin{figure}[t]
  \includegraphics[width=0.45\textwidth]{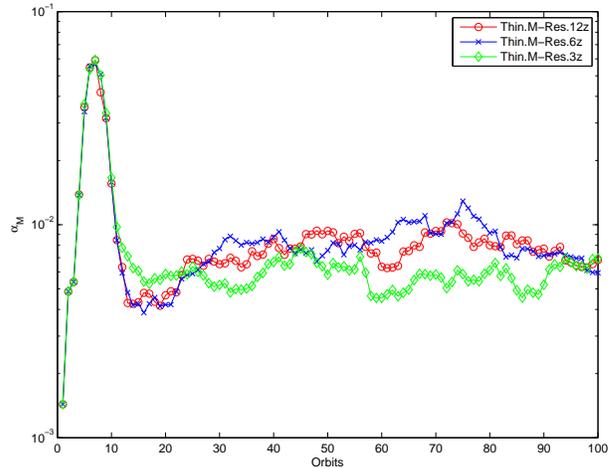}
  \caption{Behavior of $\alpha_M$ and its dependence on vertical
    domain.}
  \label{alphacomp}
\end{figure}

A comparison of $\alpha_{\rm M}$ and its dependence on vertical domain
is given in Figure~\ref{alphacomp}.  The initial growth phase of the
MRI is unaffected by the vertical domain as expected, since all the
simulations considered have the same vertical resolution and can thus
resolve the same unstable MRI modes.  Over the course of the
simulations, the larger vertical extent simulations have, in general,
larger values of $\alpha_{\rm M}$.  We attribute this effect to
stifling of the growth of the magnetized regions in the smaller
vertical extent simulation.  However, the long-term effects of
vertical extent are ambiguous.  Whether the simulations converge to
the same $\alpha_{\rm M}$ or the apparent convergence is a result of
short-term variability is unclear from the current simulations.
Longer simulations will need to be carried out to determine the
vertical domain size that is needed in order to reliably capture the
dynamics of a global disk.   

\subsection{Long term behavior}
\label{sec_33}

\begin{figure}
  \includegraphics[width=0.45\textwidth]{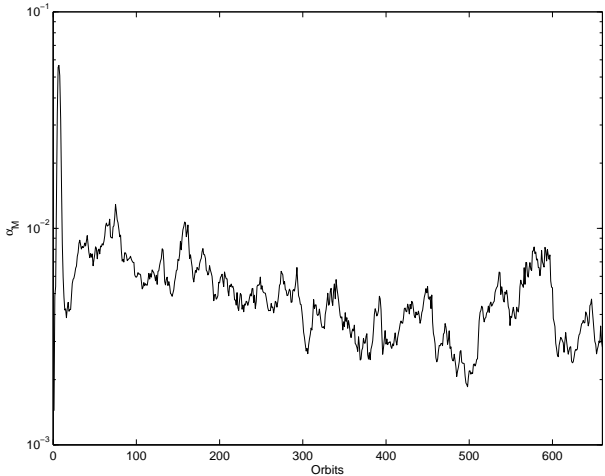}
  \caption{Long term behavior of $\alpha_M$ for run {\tt
      Thin.M-Res.6z}.}
  \label{longtermalpha}
\end{figure}

\begin{figure}
  \includegraphics[width=0.45\textwidth]{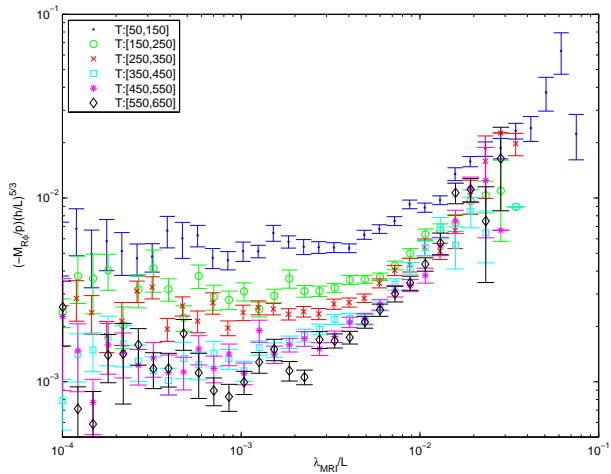}
  \caption{Flux versus stress averaged over 100 orbit blocks for run
    {\tt Thin.M-Res.6z}.  Vertical centering of $z=h$.}
  \label{longfvsblocks}
\end{figure}

In the one case of run {\tt Thin.M-Res.6z}, the disk was simulated for
664 ISCO orbits.  This simulation allows us to search for long-term
trends in the dynamics of the disk.  As shown in
Figure~\ref{longtermalpha} there is a slight downward drift in
$\alpha_M$ over time. The same temporal trend is also evident in the
flux-stress relationship. Figure~\ref{longfvsblocks} shows the
flux-stress relationship averaged in 100 orbit blocks starting at 50
ISCO orbits.  During the first 300 orbits, there appears to be a
secular drift in the flux-stress relation.  The linear (high-flux)
part of the relation achieves a steady state relatively quickly (only
the first time block between 50 and 150 ISCO orbits shows significant
differences), but the flat (low-flux) part of the relation continues
to fall until it too achieves a steady state at approximately 350 ISCO
orbits into the run.  Associated with this, the ``knee'' in the
flux-stress relation appears to move to smaller fluxes.

In essence, this result says that low-flux regions still support
(small) stresses at early times but that those stresses decay over a
period of several hundred ISCO orbits.  We ascribe this to stresses
associated with a sheared residual of the initial magnetic field
configuration which are ``mixed away'' on a relatively long timescale.
Our initial field configuration threads the midplane with regions of
net magnetic flux which alternate with a radial periodicity of $5h$.
Radial Fourier transforms of the mid-plane azimuthally-averaged $B_z$
do indeed find a (weak) periodicity corresponding to the initial field
even once the turbulence is fully developed.  This periodic component
grows weaker and is no longer detectable at approximately the same
time that the flux-stress curve achieves steady-state.  These
observations further suggest that residual flux from the initial
conditions is responsible for the long term variability.

Assuming that a long-lived residual of the initial magnetic field is
the driving mechanism for this phenomenon, we can recover the time
required to achieve the steady state from elementary arguments.  The
time needed to turbulently diffuse together two patches of oppositely
directed flux separated by a radial distance $\Delta r=2.5h$ is given
by
\begin{equation}
  t_{\rm mix}\sim \frac{\Delta r^2}{\eta_{\rm eff}},
\end{equation}
where $\eta_{\rm eff}$ is the effective turbulent resistivity.  If we
define $Pr_{\rm m,eff}$ as the effective turbulent magnetic Prandtl
number (i.e. the ratio of the effective turbulent viscosity to the
effective turbulent resistivity), we can write
\begin{equation}
  \eta_{\rm eff}=Pr_{\rm m,eff}^{-1}\alpha_{\rm M}c_sh,
\end{equation}
where $c_s$ is the sound speed.  We can then write the mixing time as
\begin{equation}
  t_{\rm mix}\sim \frac{Pr_{\rm m,eff}}{2\pi\alpha_M}\left(\frac{\Delta
      r}{h}\right)^2t_{\rm orb},
\end{equation}
where $t_{\rm orb}$ is the local orbital period and we have used the
fact that $h/c_s\sim r/v_\phi\sim t_{\rm orb}/2\pi$.  Using $Pr_{\rm
  m,eff}=1$ \citep{guan09b,lesur09,fromang09} and $\alpha_M=0.005$
suggests that the memory of the initial conditions will be lost on a
timescale of $t_{\rm mix}\sim 200t_{\rm orb}$.  This crude estimate is
in reasonable agreement with the timescale on which we see the
flux-stress relationship achieve a stationary state.

\section{Conclusions}

It has been a long-held ansatz that one can extract and model the
dynamics of a local patch of an accretion disk and obtain results (for
the angular momentum transport, for example) that have meaning for the
disk as a whole.  By examining local patches of a high resolution
global disk simulation, we have provided a direct test of this notion.
We have shown that MRI-driven turbulence in global geometrically thin
accretion disks behaves in a way consistent with scaling laws derived
for local simulations.  In particular, we find that global disks
display a local flux-stress relation qualitatively similar to that
found in local simulations \citep{hawley95,boxscaling}.  However,
other aspects of the global models are distinctly different:
\begin{enumerate}
\item Even though we model a global accretion disk that has zero net
  magnetic field, any given patch of the disk is threaded by a net
  magnetic flux resulting from the self-generated field in the
  MRI-dynamo.  Across much of the disk, the local flux is strong
  enough to have a controlling effect on the local stress.  Thus, our
  {\it zero-net field} global disk is behaving as a collection of {\it
    net field} local patches.
\item The normalization, slope, and location of the ``knee'' of the
  flux-stress relationship changes with vertical height in the
  accretion disk.  This amplifies the role of the off-midplane region
  ($h<|z|<2h$) of the disk; not only does this region have stronger
  vertical magnetic fields than the midplane, but a given vertical
  field induces stronger magnetic stresses.  The result is a strong
  enhancement of magnetic stress well off the midplane of the disk.
\item The transition point (or ``knee'') in the flux-stress relation
  occurs as significantly smaller fluxes in the global simulation as
  compared to the local unstratified simulations.   We relate this
  transition point to the ability of the simulation to marginally
  resolved the slowest appreciably growing mode.
\item Angular momentum transport (i.e. $\alpha_M$) in the global disk
  appears to be impeded if the vertical domain size of the simulation
  is too small; we found significant differences between our $z=\pm5h$
  and $z=\pm10h$ cases.  On the other hand, the $z=\pm10h$ and
  $z=\pm20h$ cases appear very similar suggesting convergence has been
  achieved.  Given that magnetic linkages between different patches of
  the disk appear to be crucial for determining the local flux (and
  hence the local stress), and that such linkages are made through the
  low-density corona of the disk, such a sensitivity to the vertical
  domain size is not surprising.
\item Analysis of our long simulation (which ran for 664 ISCO orbits)
  reveals long-term secular trends.  In particular, there is a secular
  drift of the flux-stress relationship such that the knee of the
  relationship moves to smaller fluxes and the low-flux normalization
  decreases.  This secular drift stabilizes after approximately 300
  ISCO orbits.  We attribute this to stresses associated with a
  long-lived residual of the initial magnetic field configuration.
\end{enumerate}

We would like to thank Cole Miller, Sean O'Neill, Aaron Skinner, Eve Ostriker, and
Jim Stone for valuable discussions and comments, and the Isaac Newton
Institute for Mathematical Sciences for their hospitality during the
completion of this work. K.A.S. thanks the Maryland-Goddard Joint
Space Science Institute (JSI) for support under their JSI graduate
fellowship program.  K.A.S. and C.S.R. gratefully acknowledge support
by the National Science Foundation under grant AST-0607428.
P.J.A. acknowledges support from the NSF (AST-0807471), from NASA's
Origins of Solar Systems program (NNX09AB90G), and from NASA's
Astrophysics Theory program (NNX07AH08G).

\bibliographystyle{apj}
%\bibliography{ms}

\end{document}